\documentclass[12pt]{article}

\usepackage{epsfig}

\newcommand{\gsim}{\mathrel{\vcenter
    {\hbox{$>$}\nointerlineskip\hbox{$\sim$}}}}

\newcommand{\gev}{\,{\rm GeV}}

\begin{document}

\begin{flushright}
  CPHT--S037.0901 \\
  DESY--01--137 \\
  hep-ph/0110080
\end{flushright}

\vspace{\baselineskip}

\begin{center}
\textbf{\LARGE Probing generalized parton distributions in
\\[0.5\baselineskip]
$\pi N \to \ell^+\ell^-\, N$
} \\
\vspace{3\baselineskip}
{\large
E. R. Berger\,${}^{a,}$\footnote{Email:
edgar.berger@cpht.polytechnique.fr},
M. Diehl\,${}^{b,}$\footnote{Email: markus.diehl@desy.de},
B. Pire\,${}^{a,}$\footnote{Email: bernard.pire@cpht.polytechnique.fr}
}
\\
\vspace{2\baselineskip}
${}^a$\,CPHT\footnote{Unit{\'e} mixte C7644 du CNRS}, {\'E}cole
Polytechnique, 91128 Palaiseau, France \\[0.5\baselineskip]
${}^b$\,Deutsches Elektronen-Synchroton DESY, 22603 Hamburg, Germany
\\
\textit{present address:} \\
Institut f\"ur Theoretische Physik E, RWTH Aachen, 52056 Aachen,
Germany \\
\vspace{5\baselineskip}
\textbf{Abstract}\\
\vspace{1\baselineskip}
\parbox{0.9\textwidth}{We study the exclusive reactions $\pi^- p\to
\ell^+\ell^-\, n$ and $\pi^+ n\to \ell^+\ell^-\, p$ in view of
possible future experiments with high-intensity pion beams.  For large
invariant mass of the lepton pair $\ell^+\ell^-$ and small squared
momentum transfer to the nucleon these are hard-scattering processes
providing access to generalized parton distributions.  We estimate the
cross section for these reactions, explore their connection with the
pion form factor, and discuss the role they can play in improving our
understanding of the relevant reaction mechanisms.}
\end{center}

\newpage

\section{Introduction}

Hard exclusive scattering on nucleon targets has been studied in the
last few years with the aim of extracting generalized parton
distributions, quantities that contain a wealth of information on the
quark and gluon structure of hadrons
\cite{Muller:1994fv,Ji:1997ek,Radyushkin:1997ki}.  Among the channels
most studied is the production of mesons ($\gamma^* N \to \pi N$,
$\gamma^* N \to \rho N$, \ldots) or real photons ($\gamma^* N \to
\gamma N$) induced by highly virtual spacelike photons from a lepton
beam, cf.\ \cite{Goeke:2001tz} for a recent review.  High-intensity
neutrino beam facilities under discussion \cite{Geer:1998iz} may
permit the study of charged current induced processes such as $W^{*} N
\to D_s N$ \cite{Lehmann-Dronke:2001wu}.  The secondary pion beams at
such accelerators or at standalone proton beam facilities \cite{KEK},
with energies in the range of a few 10~GeV, could at the same time be
used for experiments on hadron targets.  Here we consider the
exclusive production of a high-mass timelike photon decaying into a
lepton pair, $\pi N \to \gamma^* N \to \ell^+\ell^-\, N$ with
$\ell=e,\mu$, and argue that it would be a valuable complement to the
previously mentioned processes.  It may be seen as the analog of the
timelike Compton process, $\gamma N \to \gamma^* N \to \ell^+\ell^-\,
N$, recently investigated by us~\cite{Berger:2001}.

Two-photon processes like $\gamma^* N \to \gamma N$ or $\gamma N \to
\gamma^* N$ are in many ways complementary to their meson counterparts
such as $\gamma^* N \to \pi N$ or $\pi N \to \gamma^* N$.  The theory
description is simpler for Compton scattering, and one can expect the
hard-scattering regime to be reached at lower values of the photon
virtuality.  The analysis of meson production, on the other hand, is
simpler because there is no competition with an electromagnetic
Bethe-Heitler process.  Pions further simplify the structure of the
cross section as they are spinless.  Information beyond the Compton
channels is indispensable to disentangle the flavor and spin degrees
of freedom of the generalized parton distributions.  Finally, the
process we study here involves beams of hadrons instead of leptons or
photons, and will thus allow studies under quite different
experimental conditions.  We consider both $\pi^- p\to \ell^+\ell^-\,
n$ and $\pi^+ n\to \ell^+\ell^-\, p$, which make use of different
beams and targets and present different requirements when the outgoing
nucleon is to be detected.

\section{The scaling limit}

\begin{figure}
\begin{center}
     \epsfxsize=0.95\textwidth
     \epsffile{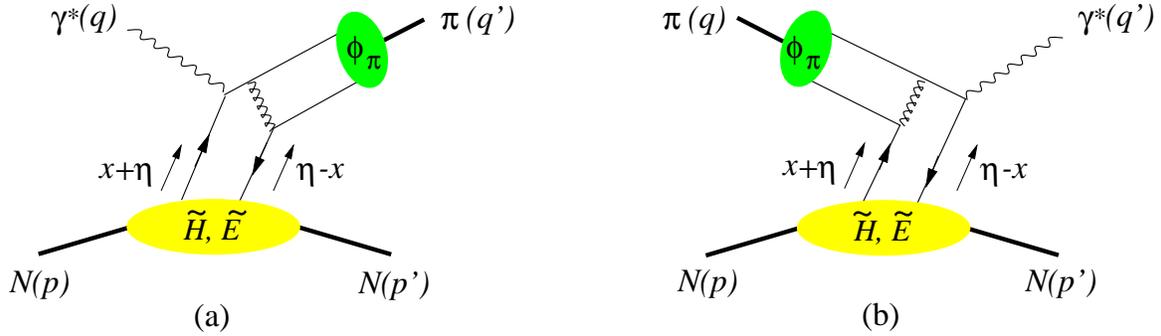}
\caption{\label{fig:hard}Sample Feynman diagrams at leading order in
$\alpha_s$ for pion electroproduction (a) and its timelike counterpart
(b) in the scaling limit.  In both cases three other diagrams are
obtained by attaching the photon to the quark lines in all possible
ways.  The plus-momentum fractions $x$ and $\eta$ refer to the average
nucleon momentum $\frac{1}{2}(p+p')$.}
\end{center}
\end{figure}

A factorization theorem \cite{Collins:1997fb} can been proven for pion
production $\gamma^* N \to \pi N$.  Its contents is represented in
Fig.~\ref{fig:hard}a, where we also define the relevant four-momenta.
In the limit of large photon virtuality $Q^2 = -q^2$ at fixed scaling
variable $x_B = Q^2 /(2 p\cdot q)$ and invariant momentum transfer $t
= (p-p')^2$ the amplitude can be written in terms of a hard-scattering
process at parton level, a distribution amplitude $\phi_\pi$
describing the formation of the pion from a $q\bar{q}$ pair, and
generalized parton distributions $\tilde{H}$ and $\tilde{E}$ encoding
nonperturbative physics in the nucleon.  The arguments for
factorization do not rely on the photon being spacelike and can be
extended to the case $\pi N \to \gamma^* N$, shown in
Fig.~\ref{fig:hard}b, with the same nonperturbative input.  The
appropriate kinematical limit is now that of large timelike virtuality
$Q'^2 = q'^2$ at fixed $t$ and fixed scaling variable
\begin{equation}
\tau = \frac{Q'^2}{2 p\cdot q}
\approx \frac{Q'^2}{s-M^2} ,
\end{equation}
where $s=(p+q)^2$ is the squared c.m.\ energy.  Here and in the
following we neglect the masses of the pion and the final-state
leptons compared with the nucleon mass $M$.

%%%%%%%%%%%%%%%%%%%%%%%%%%%%%%%%% FIGURE
%%%%%%%%%%%%%%%%%%%%%%%%%%%%%%%%%%%%

Among the predictions of the factorization theorem is that in the
limit of large virtuality the dominant polarization of the $\gamma^*$
is longitudinal in the collision c.m.  The corresponding amplitude for
$\pi N \to \gamma^* N$ scales like $1/Q'$ at fixed $t$ and $\tau$, up
to logarithmic modifications due to radiative corrections.  Transverse
photon helicity is suppressed by an extra factor of $1/Q'$ in the
amplitude.  In the limit where it can be neglected the cross section
for the overall process $\pi N \to \ell^+\ell^-\, N$ is simply
\begin{equation}
\frac{d\sigma}{dQ'^2\, dt\, d(\cos\theta)\, d\varphi}
= \frac{\alpha_{\mathrm{em}}}{256\, \pi^3}\,
  \frac{\tau^2}{Q'^6}\,
  \sum_{\lambda',\lambda} |M^{0\lambda',\lambda}|^2\,
  \sin^2\theta ,
\label{cross-section}
\end{equation}
where the superscript 0 stands for a longitudinal photon and we have
respectively taken the average and sum over the initial and final
nucleon helicities $\lambda$ and $\lambda'$.  The decay angles
$\theta$ and $\varphi$ of the photon in its rest frame are defined in
analogy to timelike Compton scattering (cf.\ Fig.~5 of
\cite{Berger:2001}), and the $\sin^2\theta$ behavior in
(\ref{cross-section}) is the sign of the purely longitudinal
$\gamma^*$ polarization.  In general the distribution in these angles
allows separation of the contributions to the cross section from
longitudinal and transverse photons, as well as their different
interference terms.  Along the lines of \cite{Diehl:1997bu} one can
thus test whether $Q'^2$ is large enough to ensure the $Q'$ behavior
and suppression pattern of the different helicity transitions
predicted by the factorization theorem.  With polarized nucleon
targets one has further access to different combinations of nucleon
helicities, in analogy with the case of $\ell N\to \ell \pi N$
\cite{Frankfurt:1999fp}.

In the large $Q'$ limit the helicity amplitudes
$M^{0\lambda',\lambda}$ for $\pi^- p \to \gamma^* n$ read
\begin{eqnarray}
\lefteqn{
M^{0\lambda',\lambda}(\pi^- p\to \gamma^* n)
 = - i e\, \frac{4\pi}{3}\, \frac{f_\pi}{Q'}\,
}
\nonumber \\
&\times&  \frac{1}{(p+p')^+}\, \bar{u}(p',\lambda') \left[
  \gamma^+\gamma_5\;  \tilde{\cal{H}}^{du}(-\eta,\eta,t) +
  \gamma_5 \frac{(p'-p)^+}{2 M}\;  \tilde{\cal{E}}^{du}(-\eta,\eta,t)
  \right] u(p,\lambda) .
\label{hel-amp}
\end{eqnarray}
An analogous equation holds for $\pi^+ n \to \gamma^* p$ with
$\tilde{\cal H}^{du}$ replaced by $\tilde{\cal H}^{ud}$ and
$\tilde{\cal E}^{du}$ by $\tilde{\cal E}^{ud}$.  Here $e$ is the
positron charge and $f_\pi \approx 132 \mbox{~MeV}$ the pion decay
constant.  We have introduced plus- and minus-components $v^\pm = (v^0
\pm v^3)/\sqrt{2}$ for any four-vector $v$ and work in a reference
frame where the average nucleon momentum $\frac{1}{2}(p+p')$ points
along the positive 3-axis.  $\eta = (p-p')^+ /(p+p')^+$ parameterizes
the plus-momentum transfer to the nucleon and has a value $\eta =
\tau/(2-\tau)$ in our kinematical limit.  To leading order in
$\alpha_S$ the convolution integrals $\tilde{\cal H}$ and $\tilde{\cal
E}$, normalized as in \cite{Belitsky:2001nq}, are given by
\begin{eqnarray}
\lefteqn{
  \tilde{\cal H}^{du}(\xi,\eta,t) = \frac{8}{3}\, \alpha_S\,
    \int_{-1}^1 dz\, \frac{\phi_\pi(z)}{1-z^2}
}
\nonumber \\
&& \times \int_{-1}^1 dx\, \left[ \frac{e_d}{\xi-x- i\epsilon} -
                          \frac{e_u}{\xi+x- i\epsilon} \right]
           [ \tilde{H}^{d}(x,\eta,t) - \tilde{H}^{u}(x,\eta,t) ]
\nonumber \\
\lefteqn{
  \tilde{\cal H}^{ud}(\xi,\eta,t) = \frac{8}{3}\, \alpha_S\,
    \int_{-1}^1 dz\, \frac{\phi_\pi(z)}{1-z^2}
}
\nonumber \\
&& \times \int_{-1}^1 dx\, \left[ \frac{e_u}{\xi-x- i\epsilon} -
                          \frac{e_d}{\xi+x- i\epsilon} \right]
           [ \tilde{H}^{u}(x,\eta,t) - \tilde{H}^{d}(x,\eta,t) ]
\label{convolution}
\end{eqnarray}
with $e_u=\frac{2}{3}$ and $e_d=-\frac{1}{3}$, and the pion
distribution amplitude normalized according to $\int_{-1}^{+1} dz\,
\phi_\pi(z) = 1$.  Analogous expressions give $\tilde{\cal E}^{du}$
and $\tilde{\cal E}^{ud}$.  Following \cite{Mankiewicz:1999aa} we have
used isospin invariance to express the generalized parton
distributions for $p\to n$ and $n\to p$ transitions in terms of the
flavor diagonal generalized $u$ and $d$ quark distributions
$\tilde{H}^u$, $\tilde{E}^u$, $\tilde{H}^d$, $\tilde{E}^d$ in the
proton, defined in \cite{Ji:1997ek}.  For ease of writing we have not
explicitly indicated the logarithmic $Q'^2$ dependence of $\tilde{\cal
H}$ and $\tilde{\cal E}$ due to the running of $\alpha_S$ and the
factorization scale dependence of the parton distributions and the
pion distribution amplitude.  The factorization scale dependence is
governed by evolution equations, which for generalized parton
distributions are a hybrid of the usual DGLAP equations for parton
densities and the ERBL equations \cite{Lepage:1979zb} for meson
distribution amplitudes, cf.\
\cite{Muller:1994fv,Ji:1997ek,Radyushkin:1997ki,Blumlein:1999sc}.

It is instructive to compare (\ref{hel-amp}) with the leading
amplitudes for the spacelike processes.  In our approximation
$M^{\lambda',0\lambda}(\gamma^* n\to \pi^- p)$ is obtained from
$M^{0\lambda',\lambda}(\pi^- p \to \gamma^* n)$ by replacing $Q'$ with
$Q$ and changing the first argument in $\tilde{\cal H}^{du}$ and in
$\tilde{\cal E}^{du}$ from $-\eta$ to $\eta$, now given by $\eta =
x_B/(2-x_B)$.  The amplitudes for $\gamma^* p\to \pi^+ n$ and $\pi^+
n\to \gamma^* p$ are connected in the same way.  We thus obtain the
simple relations
\begin{eqnarray}
M^{0\lambda',\lambda}(\pi^- p \to \gamma^* n)
  &=& \Big[ M^{\lambda',0\lambda}(\gamma^* p\to \pi^+ n)
      \Big]^* ,
\nonumber \\
M^{0\lambda',\lambda}(\pi^+ n \to \gamma^* p)
  &=& \Big[ M^{\lambda',0\lambda}(\gamma^* n\to \pi^- p)
      \Big]^*
\label{relations}
\end{eqnarray}
at leading power in $1/Q'$ and leading order in $\alpha_S$, to be
evaluated at the same $t$ and equal values of $\tau$, $x_B$ and $Q'$,
$Q$.  The corresponding unpolarized cross sections will thus be the
same for timelike and spacelike processes, up to appropriate phase
space factors.  We anticipate that the relations (\ref{relations}) no
longer hold at the level of corrections in $\alpha_S$ or in $1/Q$ and
$1/Q'$.

\section{Numerical estimates}

In order to estimate cross sections we need to specify the
nonperturbative input functions.  Among all mesons the distribution
amplitude of the pion is best constrained.  In particu\-lar the integral
needed in $\tilde{\cal H}$ and $\tilde{\cal E}$ is approximately
determined by data on $\gamma^*\gamma \to \pi$, and found close to the
value it takes for the asymptotic form $\phi_\pi(z) = \frac{3}{4}
(1-z^2)$ of the distribution amplitude under ERBL evolution.  This is
the form we will use here, the associated uncertainties (cf.\
e.g.~\cite{Diehl:2001dg}) being modest compared with others in the
process at hand.

Our knowledge of generalized parton distributions is still rather
limited.  $\tilde{H}^u$ and $\tilde{H}^d$ are known for $\eta=0$ and
$t=0$, where they become the usual spin dependent parton densities
$\Delta u(x)$ and $\Delta d(x)$, and their integral over $x$ is
related to the axial form factor $g_A(t)$ of the nucleon.  An ansatz
incorporating these constraints is
\begin{equation}
\tilde{H}^u(x,\eta,t) - \tilde{H}^d(x,\eta,t) =
 \left[\, \tilde{h}^u(x,\eta) - \tilde{h}^d(x,\eta) \,\right]  \,
  g_A(t)/ g_A(0) .
  \label{h-ansatz}
\end{equation}
We take the parameterization $g_A(t)/g_A(0) = (1 - t/M_A^2){}^{-2}$
with $M_A = 1.06$~GeV from \cite{Ahrens:1987xe}.  The functions
$\tilde{h}^u$ and $\tilde{h}^d$ are constructed from $\Delta u(x)$ and
$\Delta d(x)$ using a model prescription based on double distributions
\cite{Radyushkin:1997ki} and can be found in \cite{Berger:2001}.  For
$\Delta u(x)$ and $\Delta d(x)$ we use set A of the LO
parameterization by Gehrmann and Stirling~\cite{Gehrmann:1996ag},
retaining only the valence part since the polarized quark sea is still
poorly constrained by data.  We note that the factorization of the
$t$-dependence in (\ref{h-ansatz}) is a convenient ansatz because of
its simplicity, but must be taken with a grain of salt.

For the distributions $\tilde{E}$ we take a form given in
\cite{Penttinen:2000th}, motivated by results obtained in the chiral
soliton model of the nucleon:
\begin{equation}
\tilde{E}^u(x,\eta,t) - \tilde{E}^d(x,\eta,t) =
    \Theta(\eta - |x|)\, \frac{1}{\eta}\,
    \phi_\pi\Big( \frac{x}{\eta} \Big) \, F(t)
\label{e-ansatz}
\end{equation}
with the step function $\Theta$, the asymptotic pion distribution
amplitude given above, and
\begin{equation}
F(t) = \frac{4.4 \gev^2}{m_\pi^2-t}
  \left[ 1 - \frac{B\, (m_\pi^2-t)}{(1 - C t)^2}
  \right]
\label{pole-plus}
\end{equation}
with $B = 1.7 \gev^{-2}$ and $C = 0.5 \gev^{-2}$.
Note that (\ref{e-ansatz}) has only support for $|x| < \eta$, termed
ERBL region because of the form taken by the evolution equations
there.  In this region the generalized parton distributions describe
the emission of a $q\bar{q}$ pair from the initial nucleon as shown in
Fig.~\ref{fig:hard}.  The physics expressed in (\ref{e-ansatz}) is
that this $q\bar{q}$ pair comes from an off-shell pion.  As $t$
approaches $m_\pi^2$ the form factor (\ref{pole-plus}) is well
approximated by a pion pole form
\begin{equation}
F_{\mathrm{pole}}(t) = \frac{4 M^2 g_A(0)}{m_\pi^2 - t} ,
\end{equation}
where $g_A(0) \approx 1.25$.  For larger values the difference between
$F(t)$ and $F_{\mathrm{pole}}(t)$ may be seen as a correction to the
simple pole approximation or, in a different language, as a correction
for the off-shellness of the exchanged pion.  We note that the model
calculation in \cite{Penttinen:2000th} also found a contribution to
$\tilde{E}^u-\tilde{E}^d$ in the DGLAP region $|x| > \eta$, but this
was small compared to (\ref{e-ansatz}).

\begin{figure}
\begin{center}
     \epsfxsize=0.65\textwidth
       \epsffile{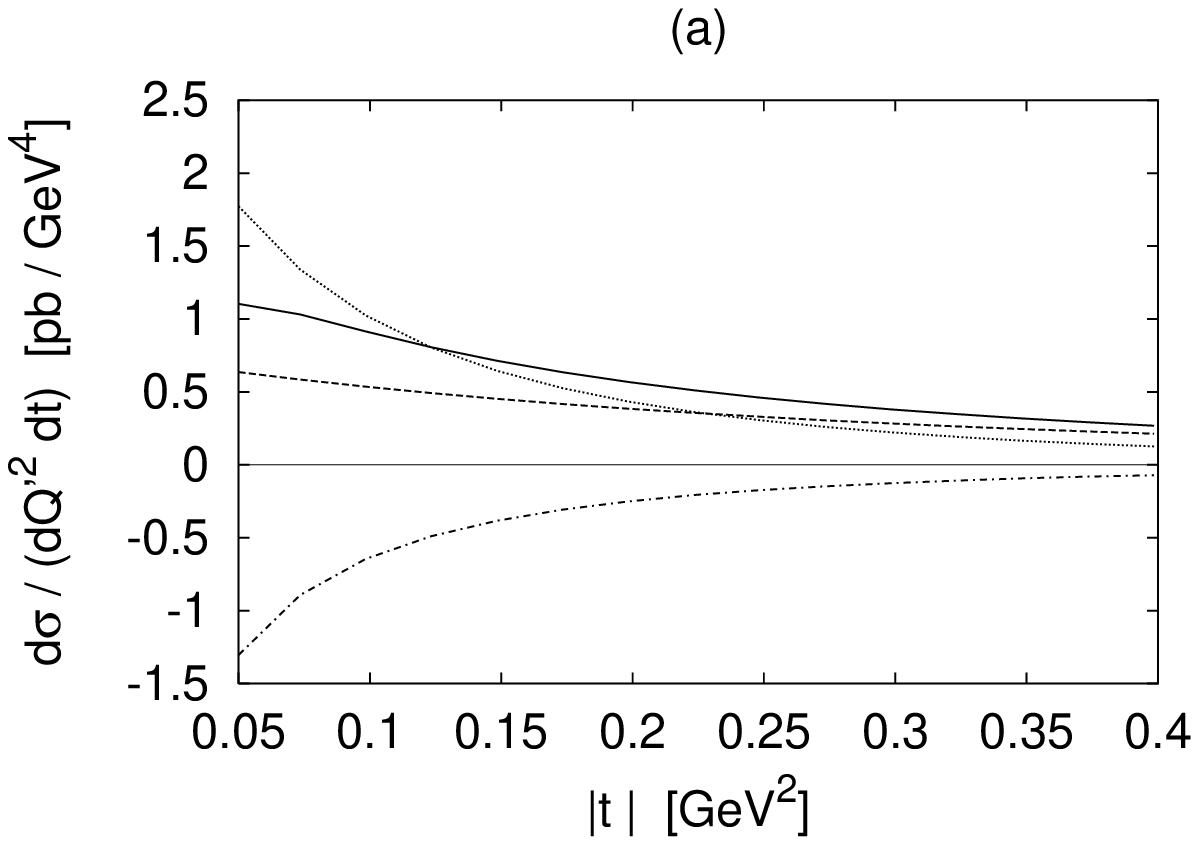}
\end{center}
\begin{center}
     \epsfxsize=0.65\textwidth
       \epsffile{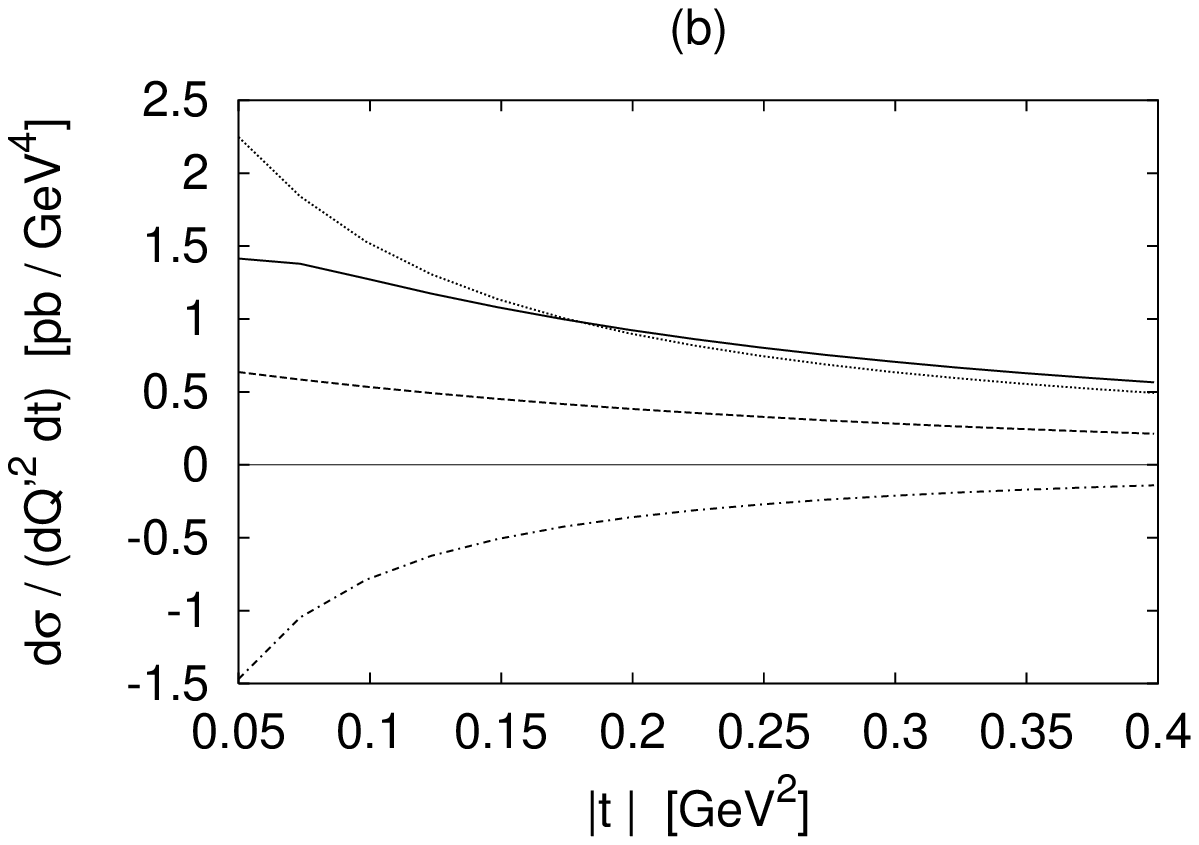}
\end{center}
\caption{\label{fig:t-plot}(a) Cross section estimates (full lines)
for $\pi^- p \to \gamma^* n$ for $Q'^2=5 \gev^2$ and $\tau = 0.2$,
calculated from (\protect\ref{full-result}) with the models
(\protect\ref{h-ansatz}) and (\protect\ref{e-ansatz}).  Separate
contributions are shown for the terms with $|\tilde{\cal H}|^2$
(dashed), $\mbox{Re} ( \tilde{\cal H}^*\, \tilde{\cal E} )$
(dash-dotted), and $|\tilde{\cal E}|^2$ (dotted). (b) The same
calculated with the pole-term form factor $F_{\mathrm{pole}}(t)$
instead of $F(t)$.}
\end{figure}

In our estimates we evaluate $\alpha_S$ at the scale $Q'^2$, using the
one-loop expression for the running coupling with 4 active flavors and
$\Lambda^{(4)} = 200 \mbox{~MeV}$ \cite{Gluck:1995uf}.  This gives
$\alpha_S \approx 0.31$ for the value of $Q'^2 = 5 \gev^2$ we consider
in the following.  In Fig.~\ref{fig:t-plot}a we show the unpolarized
cross section
\begin{eqnarray}
\lefteqn{
\frac{d\sigma}{dQ'^2\, dt}(\pi^- p \to \gamma^* n)
= \frac{4\pi \alpha_{\mathrm{em}}^2}{27}\, \frac{\tau^2}{Q'^8}\,
    f_\pi^2\,
} \hspace{4em}
\nonumber  \\
&\times&  \left[ (1-\eta^2) |\tilde{\cal H}^{du}|^2
   - 2 \eta^2\, \mbox{Re}
     \Big( \tilde{\cal H}^{du}{}^*\,
           \tilde{\cal E}^{du} \Big)
   - \eta^2 \frac{t}{4 M^2}\, |\tilde{\cal E}^{du}|^2 \,
  \right] ,
\label{full-result}
\end{eqnarray}
obtained with our models (\ref{h-ansatz}) and (\ref{e-ansatz}).  We
also plot the separate contributions from the terms with $|\tilde{\cal
H}|^2$, $\mbox{Re} ( \tilde{\cal H}^*\, \tilde{\cal E} )$, and
$|\tilde{\cal E}|^2$, and see that their relative importance is
strongly $t$ dependent.  Notice that since our ansatz for $\tilde{E}^u
- \tilde{E}^d$ vanishes in the DGLAP region, $\tilde{\cal E}$ is
purely real in the Born approximation (\ref{convolution}), in contrast
to $\tilde{\cal H}$.  Fig.~\ref{fig:t-plot}b shows the results
obtained when instead of the full form factor $F(t)$ we only take its
pole term $F_{\mathrm{pole}}(t)$.  Clearly, contributions to
$\tilde{E}$ beyond the simple pion pole are important in a wide range
of~$t$.

%%%%%%%%%%%%%%%%%%%%%%%%%%%%%%%%% FIGURE
%%%%%%%%%%%%%%%%%%%%%%%%%%%%%%%%%%%%

\begin{figure}
\begin{center}
     \epsfxsize=0.65\textwidth
       \epsffile{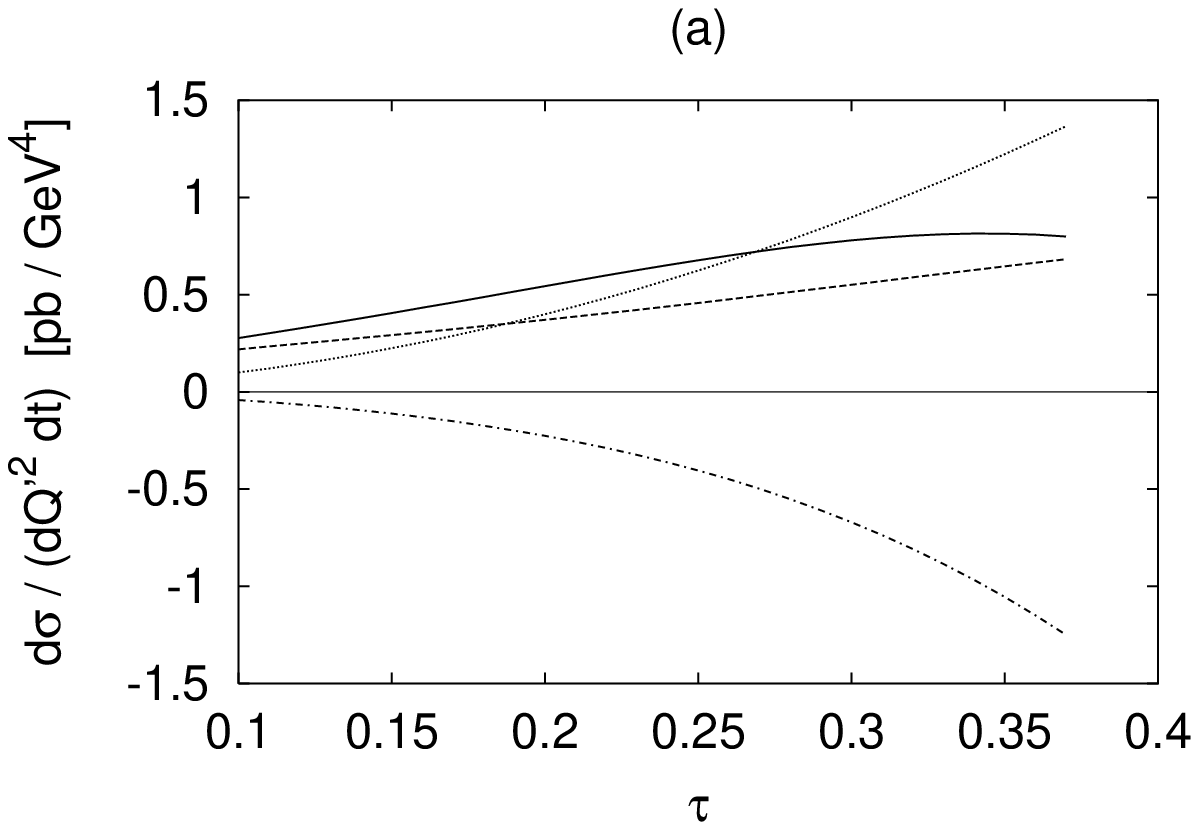}
\end{center}
\begin{center}
     \epsfxsize=0.65\textwidth
       \epsffile{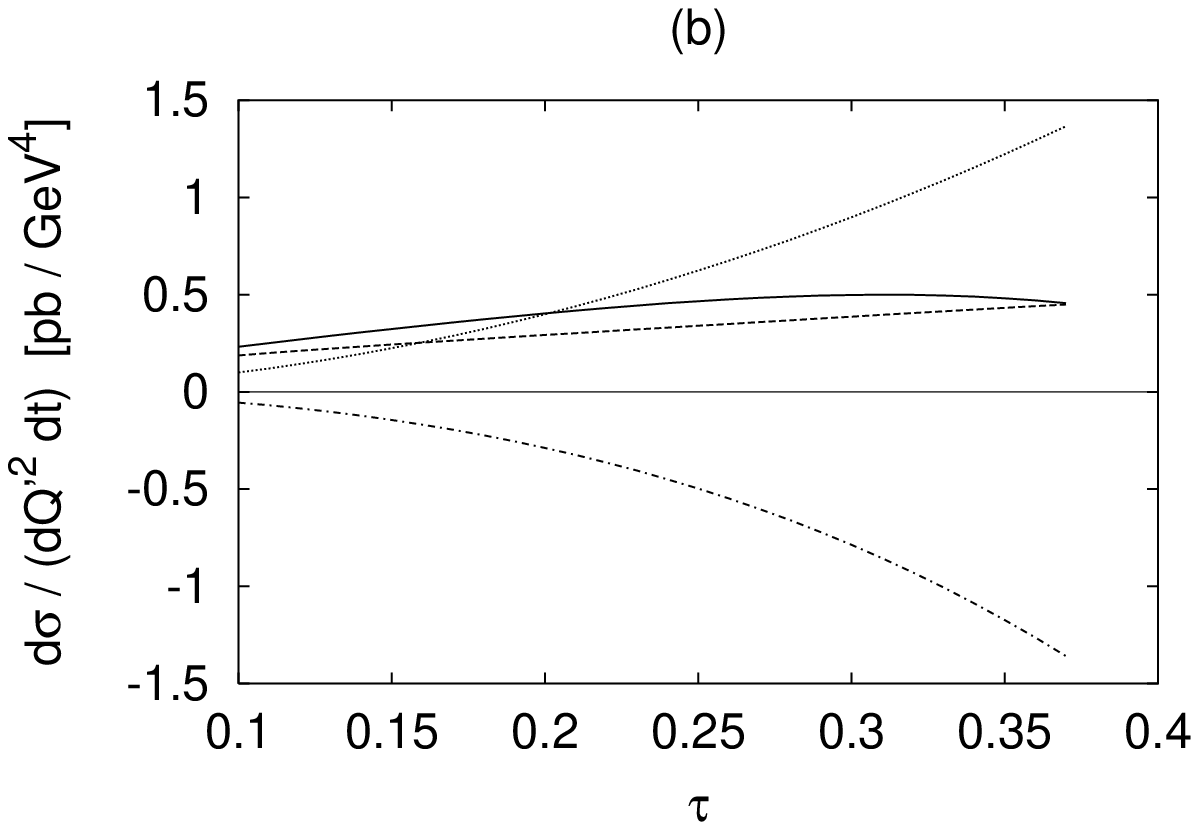}
\end{center}
\caption{\label{fig:tau-plot}(a) As Fig.~\protect\ref{fig:t-plot}a
but as a function of $\tau$ at fixed $Q'^2=5 \gev^2$ and $|t|= 0.2
\gev^2$.  (b) The same for $\pi^+ n \to \gamma^* p$.}
\end{figure}

Fig.~\ref{fig:tau-plot}a shows the cross section and its components at
fixed $Q'^2$ and $t$ as a function of $\tau$ and thus of the collision
energy $\sqrt{s}$.  We see that the relative importance of
$\tilde{\cal E}$ and $\tilde{\cal H}$ depends on both $t$ and $\tau$,
in agreement with the findings of \cite{Mankiewicz:1999kg} for the
spacelike process $\gamma^* p \to \pi^+ n$.  Where exactly one or the
other dominates depends on the details of the generalized parton
distributions.  In this respect our results should be taken with due
care, especially since we have used an oversimplified ansatz for the
$t$-dependence of $\tilde{H}$ in~(\ref{h-ansatz}).  In
Fig.~\ref{fig:tau-plot}b we show the corresponding curves for $\pi^+ n
\to \gamma^* p$.  With our ansatz (\ref{e-ansatz}) the contribution
from $\tilde{E}$ is the same for $\pi^- p$ and $\pi^+ n$, whereas the
one from $\tilde{H}$ comes out somewhat smaller for $\pi^+ n$.

%%%%%%%%%%%%%%%%%%%%%%%%%%%%%%%%% FIGURE
%%%%%%%%%%%%%%%%%%%%%%%%%%%%%%%%%%%%

\section{The pion pole}

\begin{figure}
\begin{center}
     \epsfxsize=0.75\textwidth
     \epsffile{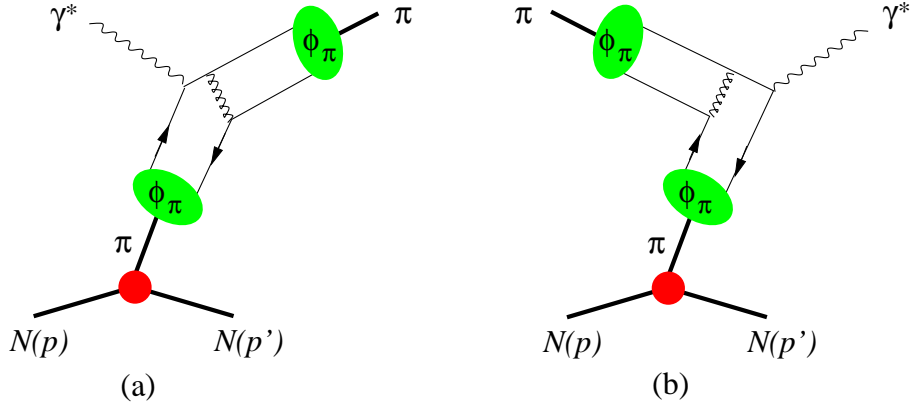}
\caption{\label{fig:pole}The part of the diagrams in
Fig.~\protect\ref{fig:hard} due to the pion pole contribution to the
distribution $\tilde{E}$.}
\end{center}
\end{figure}

We have seen that part of the amplitude in our process is due to the
pion pole.  This is represented in Fig.~\ref{fig:pole}, where the blob
representing $\tilde{E}$ in Fig.~\ref{fig:hard} has been replaced by a
nucleon-pion vertex and the pion distribution amplitude.  We recognize
in the upper part of the diagrams the space- and timelike pion form
factor $F_\pi$ in the hard-scattering formalism \cite{Lepage:1979zb}.
The connection between the pion form factors and the processes
$\gamma^* N\to \pi N$ and $\pi N\to \gamma^* N$ goes of course beyond
our approximations valid at large $Q'^2$.  For the amplitude of the
timelike process at small $t$ one may in general write
\begin{equation}
M^{0\lambda',\lambda}(\pi N\to \gamma^* N)
 = - i e\, Q' F_\pi(Q'^2)\,
    \frac{F_{\mathrm{pole}}(t)}{2 M f_\pi}\,
    \bar{u}(p',\lambda') \gamma_5\, u(p,\lambda) \,
 + \mbox{~non-pole terms} .
\label{form-factor}
\end{equation}
The ``non-pole terms'' include both off-shell corrections for pion
exchange (an example of which is the difference between $F(t)$ and
$F_{\mathrm{pole}}(t)$ discussed above) and contributions from other
sources (such as the DGLAP region of the distribution $\tilde{E}$ or
the contribution of $\tilde{H}$).

%%%%%%%%%%%%%%%%%%%%%%%%%%%%%%%%% FIGURE
%%%%%%%%%%%%%%%%%%%%%%%%%%%%%%%%%%%%

An equation analogous to (\ref{form-factor}) relates the
electroproduction process $\gamma^* N\to \pi N$ with $F_\pi(-Q^2)$,
and is in fact being used to measure the spacelike pion form factor at
large momentum transfer.  The extraction of $F_\pi(-Q^2)$ is however
not trivial because the ``non-pole terms'' in the amplitude need not
be small in the accessible kinematics and typically have to be modeled
and subtracted.  Models developed and tested for small photon
virtualities may not be adequate to describe physics at large $Q^2$
where the photon scatters not on a full off-shell pion but only on the
small-size $q\bar{q}$ component of its wave function shown in
Fig.~\ref{fig:hard} \cite{Brodsky:1998dh}.  As we have seen, even the
departure of the form factor $F(t)$ from a pure pole form is
numerically important in a wide range of $t$, as has been pointed out
earlier \cite{Carlson:1990zn}.  The timelike process presents a unique
opportunity here, since one may directly compare data for the timelike
form factor from $e^+e^- \to \pi^+\pi^-$ with data from $\pi N\to
\ell^+\ell^-\, N$ and thus \emph{test} the quality of the pion pole
approximation or of models aiming to describe corrections to it.

\section{Radiative corrections}

The ${\cal O}(\alpha_S)$ corrections to the pion form factor have been
fully calculated, cf.\ \cite{Melic:1999qr} for a recent discussion.
By a straightforward rescaling of the longitudinal momentum variables
they also give the NLO corrections to $\gamma^* N\to \pi N$
\cite{Belitsky:2001nq}.  The corresponding expressions can be
analytically continued to the timelike region; for this one needs to
replace the logarithms $\log(Q^2/\mu^2)$ by $\log(Q'^2/\mu^2) - i\pi$
in the hard-scattering kernel, where $\mu$ is either the
renormalization or the factorization scale.

Numerical studies of the spacelike form factor \cite{Melic:1999qr} and
of pion electroproduction \cite{Belitsky:2001nq} indicate that the
size of NLO corrections can be substantial and strongly depends on the
choice of renormalization scale $\mu_R$ in $\alpha_S$.  If the
asymptotic pion distribution amplitude is taken,
Ref.~\cite{Melic:1999qr} found NLO corrections to be quite small for
$\mu_R^2 \approx Q^2 /20$.  Further arguments for such a choice have
been given in \cite{Bakulev:2000uh}.  Notice that for a wide range of
$Q^2$ this requires one to modify the running of $\alpha_S(\mu_R)$ in
the infrared and thus to take into account effects beyond perturbation
theory.  This is highly nontrivial, and in our numerical studies here
we preferred to stay with the naive choice $\mu_R^2 = Q'^2$.  It is
however understood that the absolute values of our cross sections can
only be taken as rough estimates.

\section{Beyond the limit of large $Q'^2$}

In all processes we have discussed there are corrections to the
hard-scattering description of the factorization theorems as the hard
scale $Q'^2$ or $Q^2$ is not infinitely large.  An obvious issue in
reactions with timelike photons are resonance effects.  Based mainly
on the data for $e^+e^-\to {\mathit hadrons}$ we have estimated in
\cite{Berger:2001} that a description in terms of quarks and gluons
may be working for photon masses $Q'$ above, say, 1.5 or 2~GeV,
excluding the regions of the $c\bar{c}$ and the $b\bar{b}$ resonances.
We insist however that this is only a guideline since parton-hadron
duality can be realized to different degrees in different processes.
In $\pi N\to \gamma^* N$ one will further impose that $Q'$ is not too
close to the total available energy $\sqrt{s}$ in order to avoid
reinteractions between the hadronic part of the photon and the nucleon
becoming important.

We cannot give a comprehensive discussion of power corrections to the
scaling limit here, but wish to point out two distinct mechanisms that
have been studied in the literature.  One concerns contributions from
the one-gluon exchange mechanism of Figs.~\ref{fig:hard}
and~\ref{fig:pole} that go beyond the approximations giving the
leading power behavior in $1/Q$ or $1/Q'$.  They can be estimated in
the modified hard-scattering picture of \cite{Botts:1989kf} which
takes into account the relative transverse momentum $k_T$ of the
partons in the hard-scattering subprocess, or using parton-hadron
duality, for instance in the framework of light-cone sum rules.  In
the spacelike processes $\gamma^* \pi\to \pi$ and $\gamma^* N\to \pi
N$ these effects tend to decrease the scattering amplitude
\cite{Jakob:1996hd,Vanderhaeghen:1999xj,Braun:2000uj}, whereas for the
timelike form factor an enhancement over the leading term in $1/Q'$
was found \cite{Gousset:1995yh}.

\begin{figure}[th]
\begin{center}
     \epsfxsize=0.75\textwidth
     \epsffile{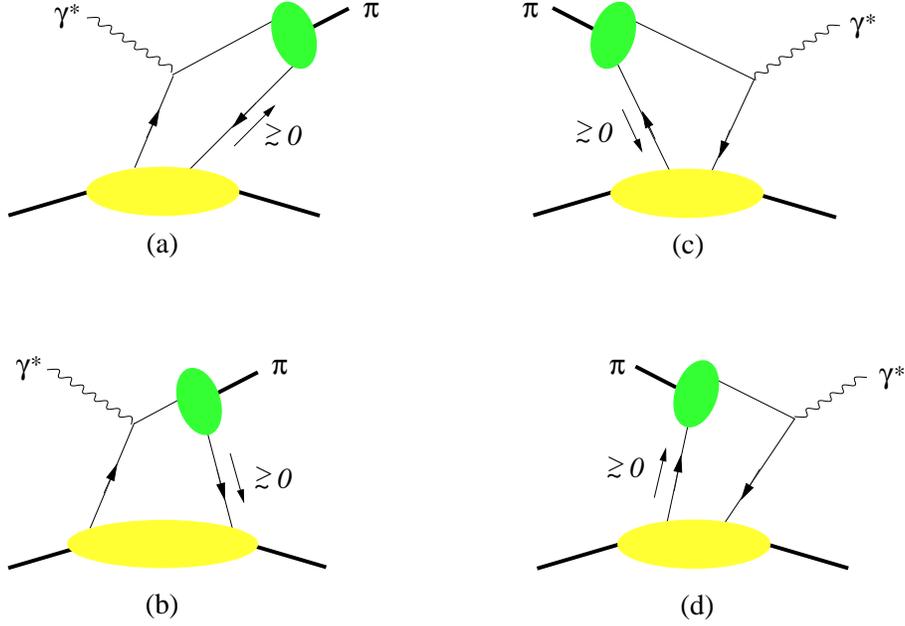}
\caption{\label{fig:soft}The soft overlap in $\gamma^* N\to \pi N$ (a
and b) and $\pi N\to \gamma^* N$ (c and d).  Plus-momenta $\gsim 0$ of
the soft partons refer to the average nucleon momentum
$\frac{1}{2}(p+p')$.  Diagrams~a, c, and d have analogs in the space-
or timelike pion form factor, with the lower blob replaced by a
quark-pion vertex.}
\end{center}
\end{figure}

A different contribution to the scattering amplitude comes from the
soft overlap mechanism, represented in Fig.~\ref{fig:soft}, where a
low-momentum parton directly connects two quark-hadron vertices.  Note
that in the case of Fig.~\ref{fig:soft}a this can be evaluated using
the formalism of light-cone wave functions
\cite{Jakob:1996hd,Vanderhaeghen:1999xj}, but not in cases b and d
where one parton in the quark-pion vertex is incoming and the other
outgoing.  Sudakov-type corrections to the quark-photon vertex in this
mechanism were investigated within parton-hadron duality
\cite{Bakulev:2000uh} and found to suppress the spacelike form factor,
whereas the timelike form factor was enhanced and also acquired an
imaginary part.  Little is known about soft overlap in the DGLAP
region of meson electroproduction, Fig.~\ref{fig:soft}b.

%%%%%%%%%%%%%%%%%%%%%%%%%%%%%%%%% FIGURE
%%%%%%%%%%%%%%%%%%%%%%%%%%%%%%%%%%%%

Many estimates find that both types of corrections just discussed can
be quite large for photon virtualities of several GeV$^2$.  In the
spacelike region they partly cancel each other, but by how much
depends on the particular model used.  The process $\pi N\to \gamma^*
N$ may help to further the understanding of these corrections.  It
allows for an unambiguous comparison of time- and spacelike data as
the latter are unaffected by the problems of extracting $F_\pi(-Q^2)$
discussed above, and the different mechanisms act differently in the
space- and timelike cases as we have seen.  One can expect that they
also show differences between the ERBL and DGLAP regions and between
pion pole and non-pole contributions, whose relative weight can be
varied by external kinematics as our above estimates suggest.

\section{Conclusions}

We argue that data on high-mass lepton pair production in the timelike
process $\pi N \to \gamma^* N$ can make important contributions to the
study of the bound-state structure of hadrons, in particular through
generalized parton distributions.  In the limit of large photon
virtuality and to leading order in $\alpha_S$, the processes $\pi N
\to \gamma^* N$ and $\gamma^* N\to \pi N$ have essentially the same
scattering amplitude, reflecting the analogous relation between the
time- and spacelike pion form factor at the same accuracy.  We have
estimated the cross section for $\pi^- p\to \ell^+\ell^-\, n$ and
$\pi^+ n\to \ell^+\ell^-\, p$ in view of possible experiments with
high-luminosity pion beams at neutrino factories or elsewhere.  We
find that the relative contributions from pion exchange and other
mechanisms strongly depend on the external kinematics, although the
details are model dependent.

Various theoretical studies as well as comparison with existing data
suggest that power corrections to the hard-scattering description of
the above processes are important, even for photon virtualities of
several GeV$^2$.  Their quantitative understanding will thus be
necessary in order to extract generalized parton distributions.  We
argue that the comparison of data on $\gamma^* N\to \pi N$ and $\pi N
\to \gamma^* N$ could offer help towards this goal, since various
correction mechanisms behave differently under the transition from
space- to timelike kinematics.  An additional handle is provided by
exploring these corrections as a function of the variables controlling
the relative importance of the DGLAP and ERBL regimes in the
generalized parton distributions.

Finally, the comparison of data on $\pi N \to \gamma^* N$ and on
$e^+e^-\to \pi \pi$ would present a unique possibility to explore pion
pole dominance in processes with highly virtual photons in a
quantitative and model independent way.  This would be of great
benefit for the endeavors to extract the spacelike pion form factor
from $\gamma^* N\to \pi N$.

%%%%%%%%%%%%%%%%%%%%%%%%%%%%%%%%%%%%%%%%%%%%%%%%%%%%%%%%%%%%%%%%%%%%%%

\end{document}